# Intelligent radiative thermostat induced by near-field radiative thermal diode

Yang Liu [a], Mauro Antezza [b, c, d], Yi Zheng [a, d, *]

[a] *Department of Mechanical and Industrial Engineering, Northeastern University, Boston, MA 02115, USA*

[b] *Laboratoire Charles Coulomb (L2C), UMR 5221 CNRS-Université de Montpellier, F-34095 Montpellier, France*

[c] *Institut Universitaire de France, 1 rue Descartes, F-75231, Paris Cedex 05, France*

[d] *Kavli Institute of Theoretical Physics (KITP), University of California, Santa Barbara CA 93106-4030, United States of America*

[*]Electronic mail: y.zheng@northeastern.edu

**ABSTRACT**

A radiative thermostat system senses its own temperature and automatically modulates heat transfer by turning on/off the cooling to maintain its temperature near a desired set point. Taking advantage of far- and near-field radiative thermal technologies, we propose an intelligent radiative thermostat induced by the combination of passive radiative cooling and near-field radiative thermal diode for thermal regulation at room temperature. The top passive radiative cooler in thermostat system with static thermal emissivity uses the cold outer space to passively cool itself all day, which can provide the bottom structure with the sub-ambient cold source. Meanwhile, using the phase-transition material vanadium dioxide, the bottom structure forms a near-field radiative thermal diode with the top cooler, which can significantly regulate the heat transfer between two terminals of the diode and then realize a stable temperature of the bottom structure. Besides, the backsided heat input of the thermostat has been taken into account according to real-world applications. Thermal performance of the proposed radiative thermostat design has been analyzed, showing that the coupling effect of static passive radiative cooling and dynamic internal heat transfer modulation can maintain an equilibrium temperature approximately locked within the phase transition region. Besides, after considering empirical indoor-to-outdoor heat flux, rendering its thermal performance closer to that of passive solar residential building walls, the calculation result proves that the radiative thermostat system can effectively modulate the temperature and stabilize it within a controllable range. Passive radiative thermostats driven by near-field radiative thermal diode can potentially enable intelligent temperature regulation technologies, for example, to moderate diurnal temperature in regions with extreme thermal swings.

## 1. Introduction

Compressor-based cooling systems consume about 20% of global electricity to provide suitable room temperatures for various buildings, which further accelerates heating effects and greenhouse gas emissions toward the environment [1]. Therefore, energy-saving cooling approaches are highly demanded for the sustainable development of the earth. Passive radiative cooling radiates energy into the cold outer space (~3 K) [2] through the atmospheric transparent window (8 μm < λ < 13 μm) and exhibit ultrahigh solar reflectance (0.3 μm < λ < 2.5 μm) without any energy input, which has become a promising approach for reducing global energy consumption [3-7]. Recently, extensive researches have achieved efficient passive daytime radiative cooling based on various materials and structures with ultrahigh solar reflectivity and infrared emissivity within the atmospheric window simultaneously, including multilayer structures [6,8-14], metamaterials [15-18], randomly distributed particle structures [19-26], and porous structures [27-31].

Although passive radiative systems can effectively reduce its surface temperature without energy consumption under high ambient temperature, its static thermal emissivity may result in overcooling in cooler climates, which potentially leads to higher heating costs. To maintain radiative cooler surfaces at temperatures moderate enough for desired residential requirement, dynamic radiative cooling using the phase-transition material vanadium dioxide ($VO_2$) has been proposed recently. $VO_2$ presents metallic or insulating state when temperature is above or below the phase transition temperature ($T_C$), respectively. The $VO_2$-based radiative cooling system will automatically turn on/off radiative cooling based on whether the temperature of cooling system is above or below $T_C$ of $VO_2$. Relevant researches of $VO_2$-based switchable radiative cooling systems were recently studied by Ono [32], Kim [33], Zhang [34], Kort-Kamp [35] and Liu [36], in which they directly utilized the visible to mid-IR absorptivity and emissivity of photonic nanostructures under direct sunlight depending on the phase transition of $VO_2$. Although current passive radiative cooling structures can cool object surfaces to sub-ambient temperatures in many practical applications, there are still thermal analysis limitations due to the ignoration of the backsided heating input or internal heat generation of objects, such as residential buildings, vehicles, and data processing centers [37]. Under practical environmental conditions, the main performance improvement with the thermal design of the passive radiative systems is how to dissipate the internal heat to the cold top surface of cooler. Therefore, designing a photonic structure that works for dynamic radiative cooling with controllable internal heat transfer channels is highly desirable and it can greatly extend the potential applications of this energy-saving cooling technology.

In this work, we propose an intelligent radiative thermostat system driven by near-field radiative thermal diode with both top-layer passive radiative cooling and internal heat transfer modulation, which can sense its own temperature and automatically maintains its temperature near a desired set point. The top radiative cooling part is a multilayer photonic structure made of polydimethylsiloxane (PDMS) and silver (Ag), acting as a static passive radiative cooler that provides the sub-ambient cold source. The bottom part is composed of $VO_2$ nanograting and gold (Au) layer, which forms a near-field radiative thermal diode with the top radiative cooler, self-adjusting near-field radiative heat transfer (NFRHT) between two terminals of the diode depending on the phase transition of $VO_2$. The simulation results show that this system can perform static top-layer radiative cooling and keep bottom-layer temperature around a desired set point, corresponding to the variable ambient temperature and constant/variable backsided heat input, without external energy consumption. Therefore, the combination of the static passive radiative cooling and near-filed radiative thermal diode can pave the way for applications of radiative thermostat within energy-efficient buildings and vehicles and energy-intelligent management systems.

## 2. Theoretical model and methods

Here, we propose a composite structure design consisting of a top passive radiative cooler made of multilayer structure of PDMS (thickness $h_1$), Ag (thickness $h_2$), and a $VO_2$-based thermal diode based on the dynamic NFRHT between the top and bottom structures depicted schematically in Fig. 1, where hexagonal boron nitride (hBN, thickness $h_3$) on the back of the Ag layer is used for NFRHT across thermal diode. The bottom structure is composed of VO$_2$ nanogratings (thickness $h_4$) and Au thin film (thickness of $h_5$) on a silicon (Si) substrate (thickness $h_6$). The VO$_2$ nanograting period and ridge width are denoted as $\Lambda$ and $w$, respectively. The nanograting filling ratio is $\phi = w/\Lambda$. These two structures are separated by a nanoscale distance $D$ much less than the thermal wavelength (0.3 $\mu$m $\leq \lambda \leq$ 20 $\mu$m). Here, we set that $h_1$ = 100 $\mu$m, $h_2$ = 500 $\mu$m, $h_3$ = 1000 nm, $h_4$ = 100 nm, $h_5$ = 1000 nm, $h_6$ = 500 $\mu$m, $\Lambda$ = 50 nm, and $D$ = 100 nm.

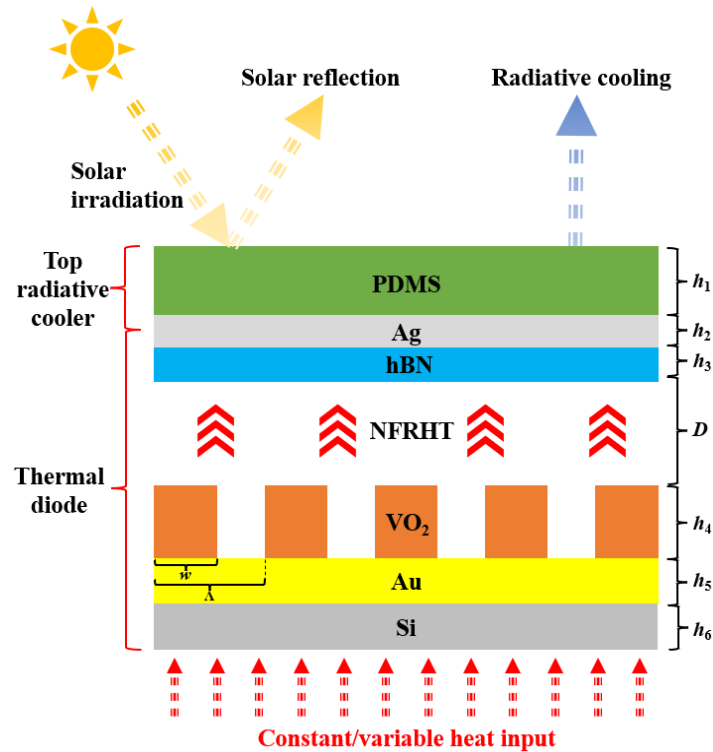

**Fig. 1.** Schematic of the intelligent radiative thermostat system induced by near-field radiative thermal diode. The top passive radiative cooler consists of multilayer thin films (PDMS and Ag). The bottom side is a one-dimensional rectangular nanograting made of VO$_2$ and Au thin film and is deposited on a Si substrate. In the calculation, a backsided constant heat input of 30 W m$^{-2}$ without sunlight was assumed.

The simulation results of spectral absorptivity, reflectivity and transmissivity of the top radiative cooler made of PDMS and Ag thin films in the spectral region (0.3 μm <

λ < 20 μm) is shown in Fig. 2. It can be observed that the radiative cooler exhibits extremely high solar reflectivity and thermal emissivity near the atmospheric window (8 μm < λ < 13 μm), because PDMS has good transparency in visible regime and ultrahigh thermal emissivity in the mid-infrared regime, which is a promising material for daytime radiative cooling [11]. Meanwhile, the solar photons transmit through the PDMS thin film and then are reflected by Ag layer, because Ag possesses excellent reflection at all bands due to its large extinction coefficient, which has been widely used as the reflecting substrate in various radiative coolers.

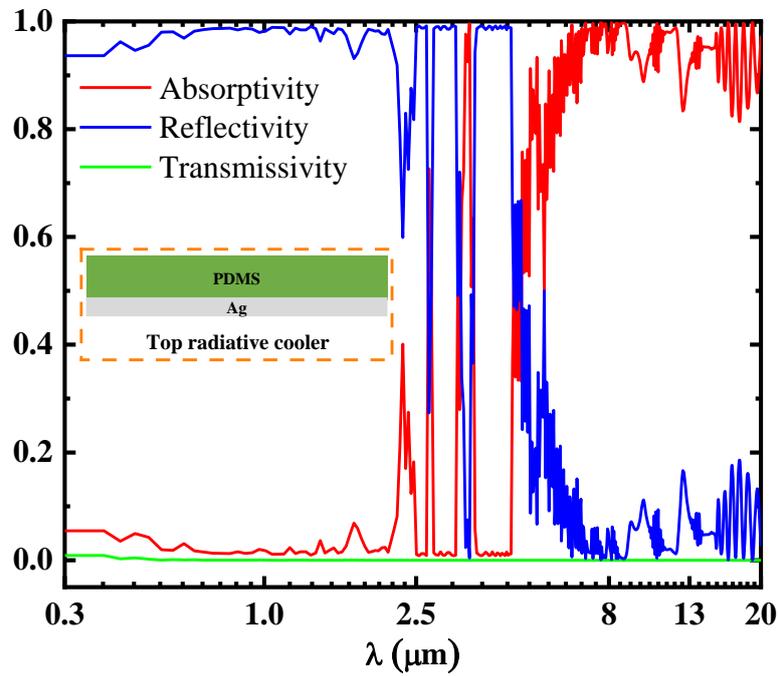

**Fig. 2.** Spectral absorptivity, reflectivity, and transmissivity of the top radiative cooler, respectively.

To calculate radiative heat fluxes across the near-field thermal diode, the expression for NFRHT obtained through the dyadic Green's function formalism is utilized as follows [38-40]:

$$Q_{1\to 2}(T_1, T_2, L) = \int_0^\infty \frac{d\omega}{2\pi} [\Theta(\omega, T_1) - \Theta(\omega, T_2)] \int_0^\infty \frac{k_\rho dk_\rho}{2\pi} \xi(\omega, k_\rho), \quad (1)$$

Here, $\Theta(\omega, T) = (\omega/2)\coth(\omega/2k_B T)$ is the energy of the harmonic oscillator. The function $\int_0^\infty \frac{k_\rho dk_\rho}{2\pi} \xi(\omega, k_\rho)$ is the spectral transmissivity of the radiative transport between two mediums separated by a gap *L*. The effective medium approximation is utilized to obtain the effective dielectric properties of the bottom $VO_2$ nanograting structure [41,42]. In addition, to evaluate the rectification efficiency of a thermal diode, we utilize the common definition of the rectification ratio $R = (Q_F - Q_R)/Q_R$, where $Q_F$ and $Q_R$ represent the forward and reverse heat fluxes between two terminals of thermal diode, respectively [41,43,44]. Meanwhile, the heat fluxes are calculated based on the temperature difference between the active and passive terminals of the diode (2Δ*T*) for the proposed $VO_2$ nanograting with different filling ratios (*ϕ* = 0.2, 0.4, 0.6, 0.8 and 1.0, respectively) shown in Fig. 1 and plotted in Fig. 3a. The temperature of the active terminal (the bottom $VO_2$ nanograting) is set as $T_1$ = 293 K + Δ*T*, while that of the passive terminal (the top radiative cooler) is $T_2$ = 293 K − Δ*T*, which is due to the phase transition temperature of $VO_2$ near 293 K [32,36]. As such, when $T_1 > T_2$ (forward bias), $VO_2$ is in the metallic phase; when $T_1 < T_2$ (reverse bias), $VO_2$ is in the insulating phase. It is evident from the Fig. 3a that the slopes of $Q_F$ are much larger than those of $Q_R$ for all structural cases, showing obvious diode-like features. Besides, when the filling ratio *ϕ* of $VO_2$ grating increases from 0.2 to 1.0, both $Q_F$ and $Q_R$ emerges the opposite trend, respectively. Figure 3b shows the rectification ratio *R* versus the filling ratio of the $VO_2$ grating. It is clearly seen that the rectification ratio *R* significantly decreases with the increasing filling ratio *ϕ*, because the filling ratio can significantly affect the optical properties of the bottom $VO_2$ grating, resulting in the changes in the surface waves across the interfaces [41]. Therefore, the filling ratio of grating can greatly affect the rectification ratio of the $VO_2$-based thermal diode.

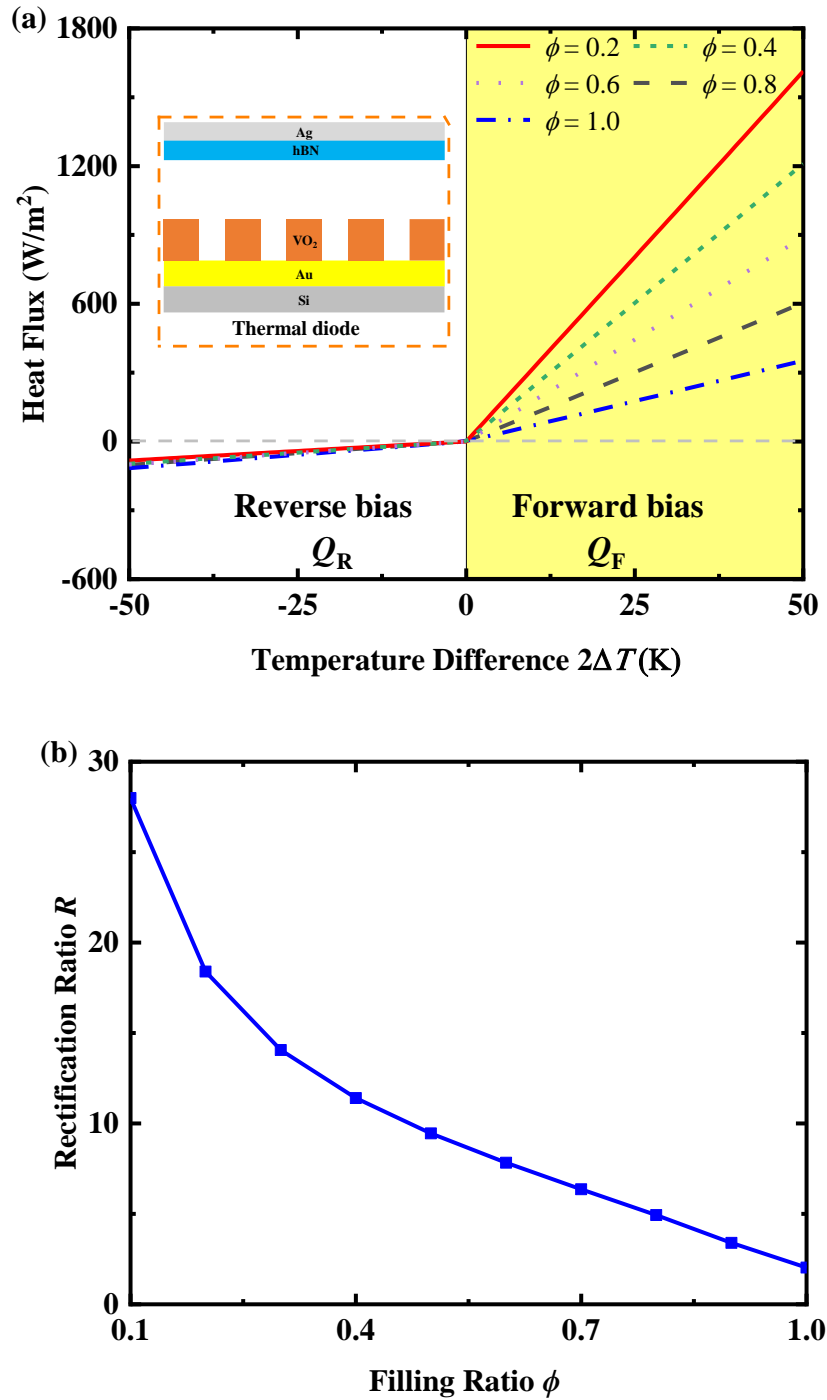

**Fig. 3.** (a) Forward and reverse heat fluxes ($Q_F$ and $Q_R$) as functions of the temperature difference between active and passive terminals of thermal diode with a 100 nm gap. (b) Rectification ratio $R$ as a function of the filling ratio of the VO$_2$ grating $\phi$.

The thermal performance of the radiative thermostat system, including the top radiative cooler and the bottom VO$_2$ grating, based on the NFRHT thermal diode is

analyzed by solving the thermal balance equation [32,45]:

$$Q_{net,cooler} = Q_{cooler}(T_{cooler}) - Q_{nr}(T_{cooler}, T_{amb}) - Q_{amb}(T_{amb}) - Q_{sun}(T_{cooler}) - Q_{diode}(T_{cooler}, T_{grating})$$

(2)

where, $Q_{net,cooler}$ and $Q_{cooler}$ are the net cooling power and radiative power of the top radiative cooler made of PDMS and Ag thin films, respectively. $Q_{nr}$ represents the non-radiative power, including heat conduction and convection. $Q_{amb}$ and $Q_{sun}$ is the incident thermal radiation from atmosphere and the incident solar energy absorbed by the top radiative cooler. $Q_{diode}$ is the heat flux between the top radiative cooler and the bottom grating structure. $T_{amb}$, $T_{cooler}$, and $T_{grating}$ are the temperatures of ambient, the top radiative cooler, and the bottom VO$_2$ grating respectively. $Q_{cooler}$ is given by [46]:

$$Q_{cooler}(T_{cooler}) = \int_0^\infty d\lambda I_{BB}(T_{cooler}, \lambda) \varepsilon(\lambda, \theta, \phi, T_{cooler})$$

(3)

In the equation, $I_{BB}(T_{cooler}, \lambda) = 2hc^2 \lambda^{-5} \left[ \exp(hc/\lambda k_B T) - 1 \right]^{-1}$ defines the spectral radiance of a blackbody at temperature $T$. The non-radiative heat transfer $Q_{nr}$ between the top radiative cooler and ambient air is given by:

$$Q_{nr}(T_{cooler}, T_{amb}) = h_{nr}(T_{amb} - T_{cooler})$$

(4)

where, $h_{nr}$ represents the non-radiative heat transfer coefficient induced by natural air heat conduction and convection [32]. Here, $h_{nr} = 8$ Wm$^{-2}$K$^{-1}$ is used throughout the calculations. $Q_{amb}(T_{amb})$ is determined as follows:

$$Q_{amb}(T_{amb}) = \int_0^\infty d\lambda I_{BB}(T_{amb}, \lambda) \varepsilon(\lambda, \theta, \phi, T_{cooler}) \varepsilon(\lambda, \theta, \phi)$$

(5)

where, $\varepsilon(\lambda, \theta, \phi)$ corresponds to the absorptivity of the atmosphere. In addition, $Q_{sun}(T_{cooler})$ represents the solar irradiation absorbed by the top radiative cooler, given by:

$$Q_{sun}(T_{cooler}) = \cos\theta_{sun} \int_0^\infty d\lambda I_{AM1.5}(\lambda) \varepsilon(\lambda, \theta_{sun}, T_{cooler})$$

(6)

where, $I_{AM1.5}(\lambda)$ is the spectral irradiance intensity of solar irradiation at AM 1.5 and $\theta_{sun}$ stands for the direction of the incoming sunlight [47]. Besides, $\varepsilon(\lambda, \theta_{sun}, T_{cooler})$ is the temperature-dependent emissivity of top radiative cooler. Meanwhile, for thermal performance of the bottom VO$_2$ grating, the net cooling power of the grating $Q_{net,grating}$

combining backsided heat input and heat transfer by thermal diode is determined as follows:

$$Q_{net,grating} = Q_{diode}(T_{cooler}, T_{grating}) - Q_{input} \qquad (7)$$

where, $Q_{input}$ indicates the heat input from the internal space to the backside of $VO_2$ grating. The time-dependent temperatures of the top radiative cooler as well as the bottom grating structure can be obtained by solving the following equations, respectively:

$$C_{cooler}\frac{dT_{cooler}}{dt} = Q_{net,cooler}(T_{cooler}, T_{amb}, T_{grating}), \qquad (8)$$

$$C_{grating}\frac{dT_{grating}}{dt} = Q_{net,grating}(T_{grating}, T_{amb}, T_{cooler}). \qquad (9)$$

Here, the heat capacitance of the radiative cooler ($C_{cooler}$) and the grating structure ($C_{grating}$) will consider the contribution of each component of the structures, as described in the literature [32,48].

## 3. Results and discussion

The thermal performance of the radiative thermostat system is evaluated by calculating the top radiative cooler and the bottom $VO_2$ grating structure ($\phi = 0.2$) temperatures, respectively, in response to the recorded 24-hour outdoor weather data. Here, we use the ambient temperature [49] and solar illumination data [50] on July 20, 2018 in Stanford, California. Figure 4 shows the temperature changes of the top radiative cooler and the bottom $VO_2$ grating over a 24-hour period, respectively. On the one hand, the temperature change of the top radiative cooler with a fixed emissivity, keeping the open-cooling state for the whole day, is shown in orange curve in Fig. 4. The top radiative cooler can yield the temperature drops at sunrise, noon, and sunset over 8 K, which does not change in response to the ambient temperature, even causing it to overcool in the cold ambient environment, such as from 12:00 AM to 9:00 AM and 06:00 PM to 12:00 AM, respectively. On the other hand, the bottom $VO_2$ grating temperature firstly drops rapidly below 293 K due to the NFRHT across the thermal diode, where the top radiative cooler provides a huge sub-ambient cold source. But due to the heat transfer inhibition induced by reverse-bias thermal diode, which restrains heat flow from the bottom grating to the top radiative cooler, the bottom grating temperature cannot drop significantly as much as that of the top cooler. Then, from 6:00 AM to 12:00 AM, the temperature of VO$_2$ grating maintains an equilibrium temperature (about 293 K) approximately locked within the phase transition region, which presents a small temperature fluctuation. Meanwhile, when the grating temperature rises over 293 K due to the backsided heat input, the thermal diode turns to the forward bias enhancing heat flow from the grating to the top cooler, which then results in the grating temperature back below 293 K. Besides, as the cooler temperature rises close to 293 K with the ambient temperature, the grating temperature will be immediately close to the cooler temperature due to the heat dissipation of forward-bias thermal diode, where the huge difference between the grating and ambient temperatures reaches a peak of about 10 K. Therefore, the temperature change of the bottom grating over this 24-hour period clearly exhibits the radiative thermostat performance, resilient to the surrounding environment.

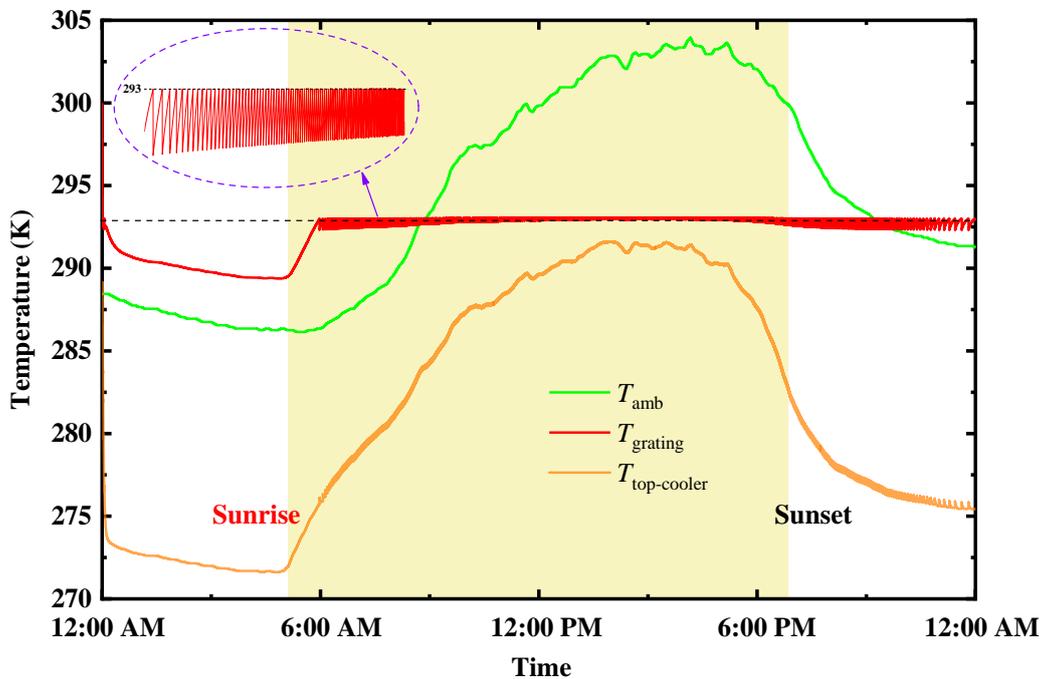

**Fig. 4.** Thermal performance of the intelligent radiative thermostat system, including the bottom grating structure (red curve) and the top radiative cooler (orange curve) over a 24-hour cycle with ambient temperature variation (green curve).

The transient temperature fluctuations of the radiative thermostat system and the hypothetical integrated radiative cooler without internal NFRHT under a fixed ambient temperature condition $T_{amb}$ = 293 K are calculated by solving Eqs. (8) and (9), which are integrated to obtain the temperature evolutions of radiative thermostat and hypothetical cooler as a function of time, as shown in Fig. 5. The temperature of the top radiative cooler in thermostat reduces and eventually reaches a thermal equilibrium temperature with a small temperature fluctuation that is about 16 K below the ambient temperature because of the top-layer passive radiative cooling effect (orange solid curve in Fig. 5). However, the temperature of the bottom grating in thermostat system firstly reduces about 1 K below the ambient temperature due to NFRHT between two terminals of thermal diode, where the thermal diode is in forward bias. Once the grating has been cooled down below $T_C$ = 293 K, the thermal diode turns to reverse bias resulting in restraining heat flow from the grating to the top cooler. Then the continuous backsided heat input and the reduced heat loss driven by the thermal diode will heat the bottom grating back to temperatures above 293 K. Therefore, due to the combined effects of top-layer radiative cooling and NFRHT across the thermal diode, the temperature of $VO_2$ grating will present a small temperature fluctuation around 293 K. As a comparison, a hypothetical integrated radiative cooler is assumed without internal NFRHT, which processes the same materials, structural parameters and outer thermal conditions as the proposed radiative thermostat. It is clearly seen from Fig. 5 that its temperature (black line) reduces and eventually reaches a thermal equilibrium temperature that is about 16 K below the ambient temperature without any fluctuation. Meanwhile, the inset figure demonstrates time-dependent temperature evolution of the top radiative cooler and the bottom grating of thermostat system within one fluctuation period. When the temperature changes of the top cooler and the bottom grating emerge the opposite trends, further indicating the dynamic thermal modulation effects of thermal diode. When the grating temperature drops rapidly below 293 K due to thermal diode enhancing heat dissipation, the cooler temperature will rise rapidly by a certain percentage, because of the different specific heat capacities of these two parts. On the contrary, when the grating temperature is below 293 K, thermal diode suppresses heat dissipation and then the grating temperature will slowly increase based on backsided heat input. Meanwhile, the top cooler temperature will slowly decrease due to passive radiative cooling effect. Therefore, to compare with a hypothetical integrated passive radiative cooler, the internal thermal diode plays a significant role in radiative thermostat performance.

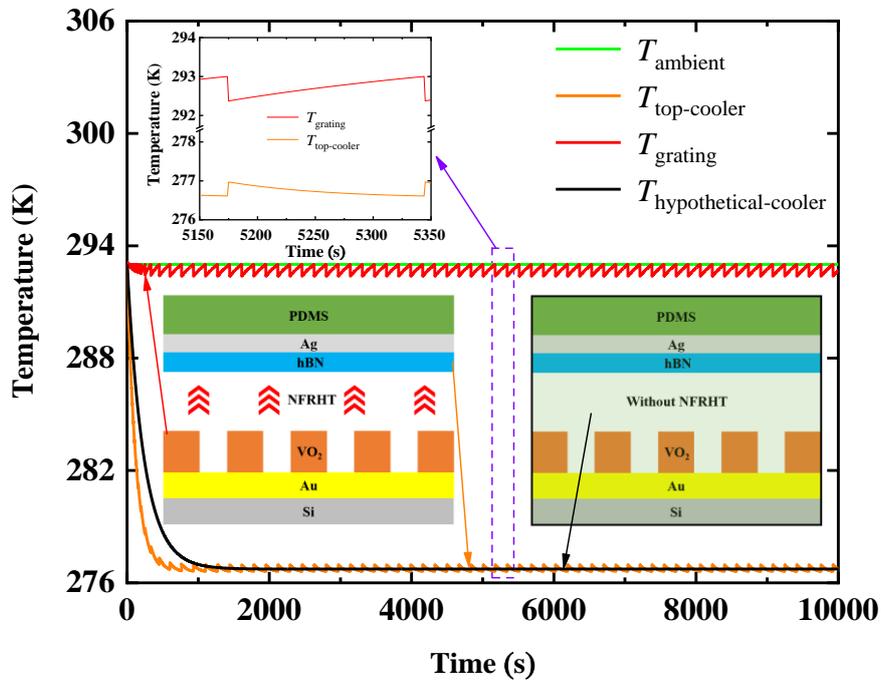

**Fig. 5.** Time-dependent temperature evolutions of the radiative thermostat (the top radiative cooler and the bottom grating) and the hypothetical integrated radiative cooler without internal NFRHT. The initial temperature of system is assumed to be the same as ambient temperature $T_{amb}$ = 293 K. The whole system is exposed to the sky at time = 0.

To further evaluate the thermal performance of our proposed radiative thermostat in passive solar residential building walls, we simulate different modes of the inner surface heat flux to heat the backside of thermostat based on the indoor temperature change and the heat capacity stored by the inner part of the wall [51,52], as shown in Fig. 6a. Mode 1 is set to be a constant heat flux of 30 W m$^{-2}$, and Mode 2 utilizes an empirical indoor-to-outdoor heat flux in a 24-hour cycle rendering the thermal performance of our proposed system closer to that of passive solar residential building walls. Furthermore, the thermal stability of the radiative thermostat system with different heat input is explored by calculating the bottom grating structure temperature in response to the recorded 24-hour outdoor weather data mentioned above, as shown in Fig. 6b. It is clearly seen that although the step-like heat input (Mode 2) causes the grating temperature to drop to some extent, where the grating temperature decreases to sub-ambient temperatures due to the smaller backsided heat input and larger diode-based heat dissipation at night-time, its temperature remains around 293 K most of the time within a controllable range (less than 1 K), further indicating the good thermal stability of the designed thermostat system.

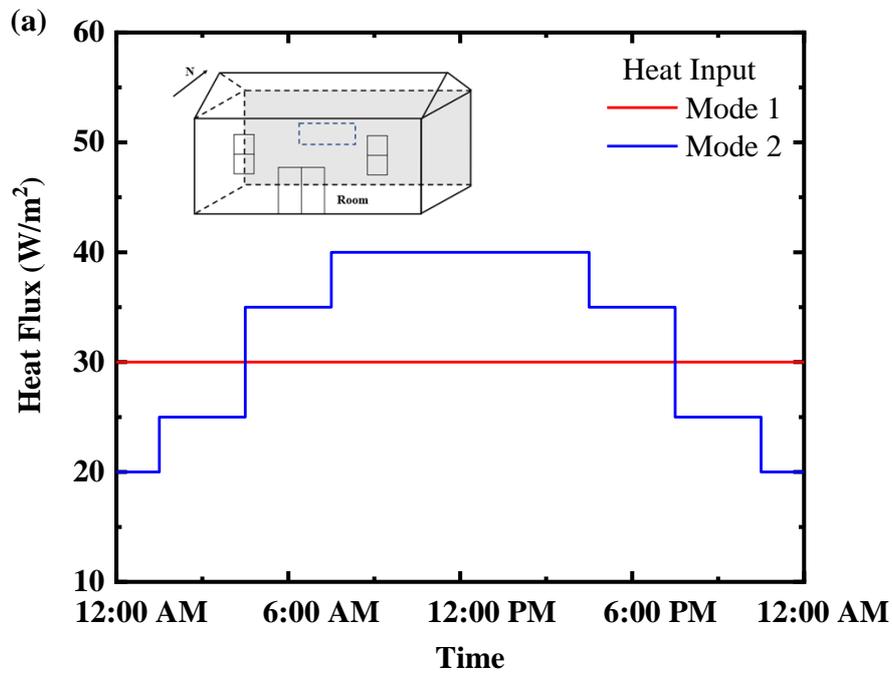

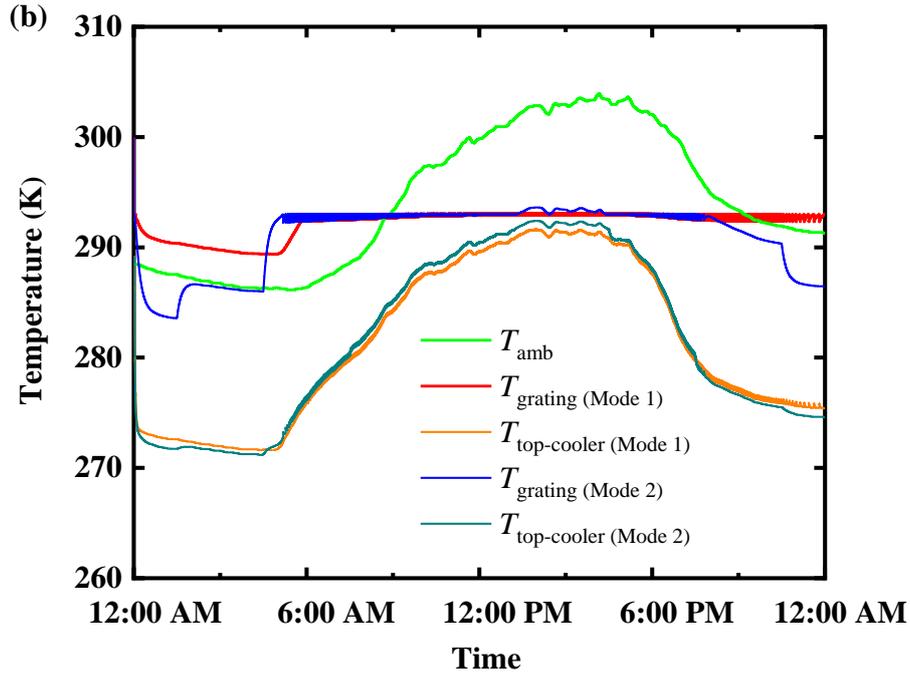

**Fig. 6.** (a) Different modes of the inner surface heat fluxes to heat the backside of thermostat. Mode 1: constant heat flux; Mode 2: empirical indoor-to-outdoor heat flux. (b) Thermal performance of the bottom grating structure with different modes of backsided heat input over a 24-hour cycle with ambient temperature variation.

## 4. Conclusions

In summary, we present an intelligent radiative thermostat system driven by near-field radiative thermal diode for near-room-temperature thermal moderation. Our radiative thermostat system can moderate temperature swings with respect to passive radiative cooler with fixed thermal emissivity, resulting in both daytime and nighttime cooling in hot climates and keeping warm in cold ones at a pre-set phase transition temperature. The top radiative cooler in thermostat system can achieve passive radiative cooling based on its own thermal emissivity, which can provide the bottom structure with sub-ambient cold source. The bottom structure of the thermostat can moderate its temperature swings with the phase transition temperature, which forms a near-field radiative thermal diode with the top cooler, realizing the dynamic heat transfer modulation between the bottom structure and the top cooler. We also simulate the empirical indoor-to-outdoor heat flux, rendering its thermal performance closer to that of passive solar residential building walls, and demonstrate that the proposed intelligent thermostat system can effectively modulate the temperature and stabilize it within a controllable range. Therefore, our designed intelligent radiative thermostat has the potential to greatly improve energy savings worldwide.

**Data availability statement**

The data that support the findings of this study are available from the corresponding author upon reasonable request.

**Declaration of competing interest**

The authors declare that they have no known competing financial interests or personal relationships that could have appeared to influence the work reported in this paper.

**Acknowledgment**

This work was supported in part by the National Science Foundation under Grant No. CBET-1941743. M.A. and Y.Z. acknowledge the KITP program 'Emerging Regimes and Implications of Quantum and Thermal Fluctuational Electrodynamics' 2022, where part of this work has been done. This research was supported in part by the National Science Foundation under Grant No. PHY-1748958.